\documentclass[12pt,preprint]{aastex}

\newcommand{\sqcm}{${\rm cm^{-2}}$}

\newcommand{\mum}{${\rm \mu m}$}
\newcommand{\kms}{${\rm km~s^{-1}}$} 
\newcommand{\waven}{${\rm cm^{-1}}$}
\newcommand{\thirteenco}{${\rm ^{13}CO}$}
\newcommand{\twelveco}{${\rm ^{12}CO}$}
\newcommand{\trot}{$T_{\rm rot}$}
\newcommand{\eighteenco}{${\rm C^{18}O}$}
\newcommand{\bdop}{$b_{\rm D}$}

\newcommand{\eqw}{$W_{\nu}$}

\usepackage{emulateapj5}
\usepackage{epsfig}

\slugcomment{To be published in ApJ (submitted 5 Oct 2001; accepted 3 Dec 2001) }

\shorttitle{High Resolution 4.7 \mum\ Spectroscopy of L1489 IRS}
\shortauthors{Boogert, Hogerheijde, \& Blake}

\begin{document}

\title{High Resolution 4.7 \mum\ Keck/NIRSPEC Spectra of Protostars.
I: Ices and Infalling Gas in the Disk of L1489
IRS\footnotemark}\footnotetext{The data presented herein were
obtained at the W.M. Keck Observatory, which is operated as a
scientific partnership among the California Institute of Technology,
the University of California and the National Aeronautics and Space
Administration.  The Observatory was made possible by the generous
financial support of the W.M. Keck Foundation.}

\author{A.C.A. Boogert\altaffilmark{2}, 
        M.R. Hogerheijde\altaffilmark{3,4},
        G.A. Blake \altaffilmark{5}}

\altaffiltext{2}{California Institute of Technology,
              Downs Laboratory of Physics 320-47, Pasadena, CA 91125,
              USA; boogert@submm.caltech.edu}
\altaffiltext{3}{RAL, University of California at Berkeley, Astronomy
                 Department, 601 Campbell Hall \# 3411, Berkeley, CA
                 94720, USA}
\altaffiltext{4}{current address: Steward Observatory, University of
                 Arizona, 933 N. Cherry Ave., Tucson, AZ 85721, USA}
\altaffiltext{5}{California Institute of Technology, Division of
              Geological and Planetary Sciences 150-21, Pasadena, CA
              91125, USA}

\begin{abstract}
 We explore the infrared M band (4.7 \mum) spectrum of the class I
protostar L1489 IRS in the Taurus Molecular Cloud.  This is the
highest resolution wide coverage spectrum at this wavelength of a low
mass protostar observed to date ($R=$25,000; $\Delta v=$12 \kms).  A
large number of narrow absorption lines of gas phase \twelveco,
\thirteenco, and \eighteenco\ are detected, as well as a prominent
band of solid \twelveco. The gas phase \twelveco\ lines have red
shifted absorption wings (up to 100 \kms), which likely originate from
warm disk material falling toward the central object.  Both the
isotopes and the extent of the \twelveco\ line wings are successfully
fitted with a contracting disk model of this evolutionary transitional
object \citep{hoge01}. This shows that the inward motions seen in
millimeter wave emission lines continue to within $\sim$0.1 AU from
the star. The amount of high velocity infalling gas is however
overestimated by this model, suggesting that only part of the disk is
infalling, e.g. a hot surface layer or hot gas in the magnetic field
tubes. The colder parts of the disk are traced by the prominent CO ice
band.  The band profile results from CO in 'polar' ices (CO mixed with
H$_2$O), and CO in 'apolar' ices.  At the high spectral resolution,
the 'apolar' component is, for the first time, resolved into two
distinct components, likely due to pure CO and CO mixed with CO$_2$,
O$_2$ and/or N$_2$.  The ices have probably experienced thermal
processing in the upper disk layer traced by our pencil absorption
beam: much of the volatile 'apolar' ices has evaporated, the depletion
factor of CO onto grains is remarkably low ($\sim$7\%), and the CO$_2$
traced in the CO band profile was possibly formed energetically.  This
study shows that high spectral resolution 4.7 \mum\ observations
provide important and unique information on the dynamics and structure
of protostellar disks and the origin and evolution of ices in these
disks.
\end{abstract}


\section{Introduction}~\label{sel1489:intro}

In the process of low mass star formation, a mixture of gas, dust, and
ices accumulates in protostellar envelopes and disks. The fate of this
molecular material is diverse. Most of it will fall toward the
protostar and dissociates in the inner disk region or stellar
photosphere. Some material will be blown away and destroyed by the
stellar wind.  Some may also survive and be the building material for
comets and planets. Major aspects of this complicated process are not
well understood, and poorly observationally constrained. For example,
do the ices that form comets still resemble ices of the original
pristine molecular clouds or are new ices of different composition
being formed in the envelope or disk? The type of ices being formed
depends on the composition of the gas that accretes onto
grains. Reducing environments produce H$_2$O-rich (`polar') ices,
while in cold inert environments `apolar' ices rich in CO, N$_2$ and
O$_2$ can be formed \citep{tiel82}.  Depending on the composition,
ices evaporate between temperatures of 18 and 90 K.  Also, heat can
change the solid state structure of ices by for example
crystallization. Energetic particles (e.g. cosmic rays) and
ultraviolet (UV) radiation are able to initiate reactions in ices and
form new species. Dynamics and shocks within disks may be able to
destroy ices as well.

Clearly, to determine the relative importance of these ice formation
and destruction processes, knowledge of the physical conditions and
structure of envelopes and disks is crucial.  Much theoretical and
observational work on this topic has been done over the last $\sim$10
years. Molecular gas was detected in a suite of protostellar disks by
millimeter wave observations sensitive to emission over radii of
several hundred AU (Dutrey, Guilloteau, \& Guelin 1997;
\citealt{thi01}).  Gas phase abundances were found to be reduced by
factors of 5 to several 100, depending on the source and the
sublimation temperature of the molecules.  Models of disk mid-planes
indeed show high depletions because of the formation of icy mantles on
grains \citep{aika97, will98}. The predicted depletions were in fact
higher than observed and thus desorption mechanisms are needed to
explain the millimeter wave observations \citep{gold99}.  It was
realized that the outer parts of disks are heated more efficiently
when they are flared \citep{keny87}. Thus, by the influence of the
stellar radiation a layer with `super-heated' dust is formed in which
molecules have been dissociated \citep{chia97}.  The layer below that
is warm enough to evaporate the ices, but not dissociate the released
molecules. The importance of this warm layer, and relative gas phase
molecular abundances, depends strongly on how effective ice desorption
mechanisms are \citep{will98}.  Recent studies indicate that
desorption by UV and X-ray photons may be strong enough to explain
observations of molecular gas by millimeter wave telescopes
\citep{will00, naji01}.  This idea is confirmed by multi transition
molecular line observations which indicate temperatures ($>20-40$ K)
and densities that typically occur in this warm layer \citep{zade01}.
In a third layer, the disk mid-plane, cold, dense conditions prevail,
resulting in extreme depletions of gas phase molecules many orders of
magnitude larger than in quiescent dense molecular clouds.  Indeed,
recent absorption line observations failed to detect gas phase CO in
the edge-on disk around the protostar Elias 18, thus indicating an
enormous depletion in the mid-plane \citep{shup01}.

In this Paper we report high spectral resolution ($R=$25,000) 4.7
\mum\ M band observations of the obscured protostar L1489 IRS (IRAS
04016+2610) in the Taurus Molecular Cloud.  L1489 IRS is a low
luminosity object (3.7 $L_{\odot}$), with a spectral energy
distribution resembling that of an embedded class I protostar (Kenyon,
Calvet, \& Hartmann 1987).  Detailed millimeter wave line and
continuum studies show that L1489 IRS is surrounded by a large, 2000
AU radius rotating thick disk-like structure \citep{hoge98, sait01},
rather than an inside-out collapsing envelope \citep{hoge00a}. The
rotation is sub-Keplerian, and the disk as a whole is
contracting. Thus it was suggested that L1489 IRS represents a
short-lasting (2$\times 10^4$ yr) transitional phase between embedded
YSOs that have large envelopes and small (few hundred AU) rotationally
supported disks, and T Tauri stars which have no envelopes and fully
rotationally supported 500-800 AU size disks \citep{hoge01}. This
circumstellar (or circumbinary: \citealt{wood01}) disk is seen close
to edge-on (60 to $<$90$^{\rm o}$) in scattered light images
\citep{whit97, padg99}. A CO outflow emanates from the object
\citep{myer88} with Herbig Haro objects lying along it \citep{gome97}.

Low spectral resolution infrared observations of L1489 IRS show that
deep H$_2$O \citep{sato90} and CO (\citealt{chia98}; Teixeira,
Emerson, \& Palumbo 1998) ice bands are present along the line of
sight. In this Paper we will use the newly available spectrometer
NIRSPEC at the Keck II telescope to obtain high resolution M band
spectra ($\Delta v=$12 \kms\ at 4.7 \mum) of this source.  The large
array of NIRSPEC allows both the vibrational band of solid CO and the
surrounding ro-vibrational transitions of gas phase CO to be observed
in the same high resolution spectrum.  This offers a new view on this
system, both on the origin and evolution of ices and the
interrelationship of gas and ices as well as on the kinematics and
structure of the young, contracting, close to edge-on disk. It is a
unique view, because infrared absorption line studies trace all gas
and solid state material at all radii from the star, while present day
millimeter wave observations are limited by their relatively low
spatial resolution ($\geq$100 AU). Thus one of the questions that will
be answered in this Paper is whether the large scale inward motions
seen in millimeter wave emission lines continue to smaller radii (1 AU
or less) from the star.

Previous studies have already shown that rich astrophysical
information can be obtained from high spectral resolution observations
in the atmospheric M band. Mainly massive, luminous protostars were
observed, however \citep{mitc90}, and the few observations of low mass
protostars cover a small wavelength range containing only a few gas
phase lines and not the solid CO band \citep{shup01, carr01}. This is
Paper I in a series on high resolution M band spectroscopy of
protostars, initiated by the availability of the NIRSPEC spectrometer
at Keck II with which weak, low mass protostars can be routinely
observed at high spectral resolution over a large wavelength range
covering both the solid and gas phase CO features.

The reduction of the long slit spectra is discussed in \S 2. In \S
3.1, we analyze the observed gas lines, which at this resolution even
give dynamical information. We use standard curve of growth and
rotation diagram techniques to get a first idea of gas column and
temperatures. In order to analyze the solid CO band profile in this
line of sight, a detailed discussion of available laboratory
experiments of solid CO is given in \S 3.2. In \S 4.1 we apply the
infalling disk model of \citet{hoge01} to explain the observed gas
phase \twelveco\ and \thirteenco\ line profiles, and constrain the
physical conditions and structure of the disk.  The possibility of
binarity is briefly discussed in \S 4.2.  The gas phase analysis is
linked to the the solid CO results to determine the origin and thermal
history of solid CO in \S 4.3. We conclude with suggestions for future
work in \S 5.

\section{Observations}

The infrared source L1489 IRS was observed with the NIRSPEC
spectrometer \citep{mcle98} at the Keck II telescope atop Mauna Kea on
the nights of 28/29 and 29/30 January 2001.  The sky was constantly
clear and dry, and the seeing was reasonable ($\sim 0.5-0.7''$ at 2.2
\mum).  NIRSPEC was used in the echelle mode with the 0.43$\times
24''$ slit, providing a resolving power of $R=\lambda/\Delta
\lambda=25,000$ ($\sim$12 \kms)\footnote{This nominal spectral
resolution was verified by measuring the Gaussian width of absorption
lines in a variety of astrophysical sources.} with three Nyquist
sampled settings covering the wavelength range 4.615--4.819 \mum\ in
the atmospheric M transmission band.

The data was reduced in a standard way, using IDL routines. The
thermal background emission was removed by differencing the nodding
pair. Each nodding position was integrated on for 1 minute, before
pointing the telescope to the other nodding position. The positioning
had to be done by hand because condensations on the dewar window
prevented us from using the image rotator, and the instrument had to
be used in the non-standard `stationary guiding mode'. In this mode,
the condensations were well removed by subtracting nodding pairs,
although the sensitivity of Keck/NIRSPEC was reduced by several
magnitudes.

\begin{figure*}[t!]
\center
\includegraphics[angle=90, scale=0.80]{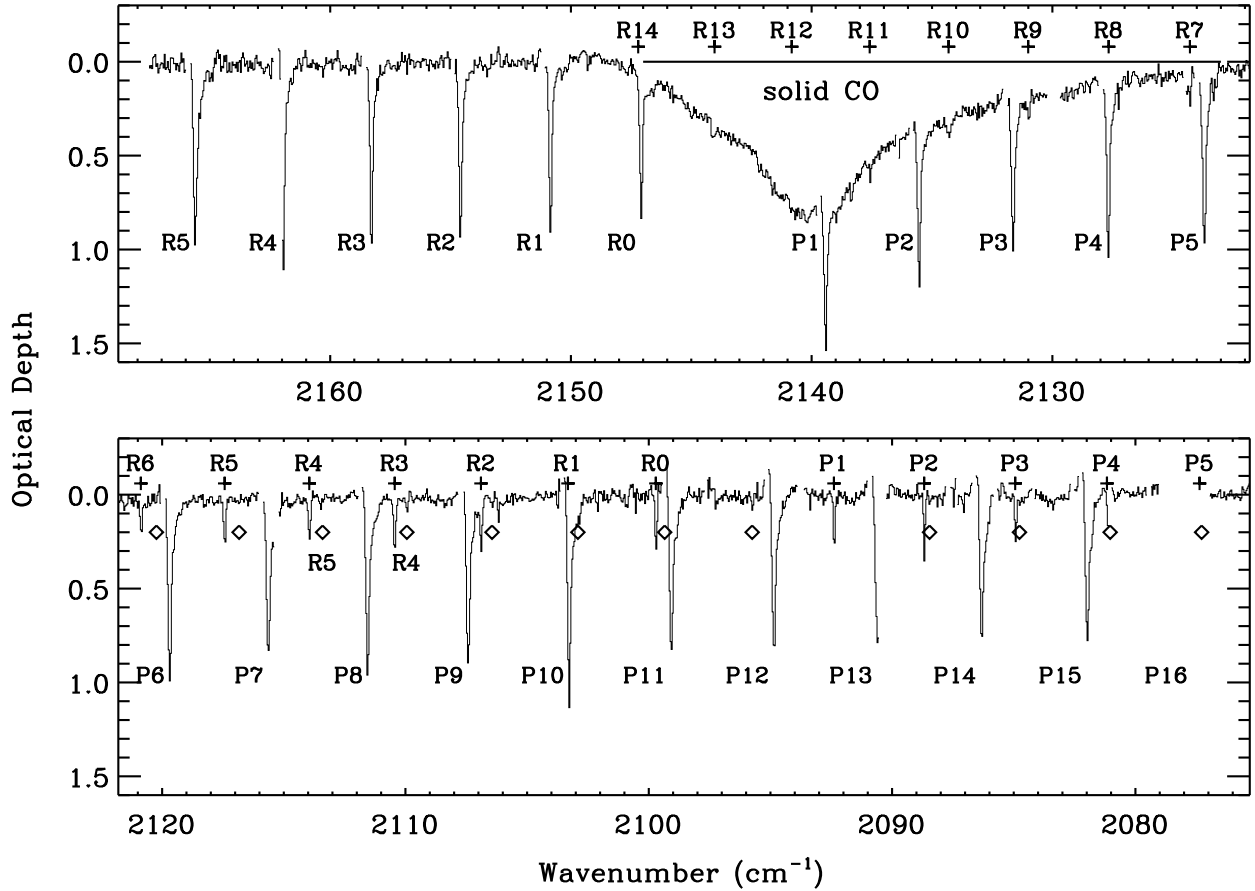}
\caption{Observed, unsmoothed, R=25,000 M band spectrum of L1489 IRS
corrected for object and earth velocity (43 \kms) on optical depth
scale, with gas phase line identifications of \twelveco\ (deepest
lines) and \thirteenco\ ('+' symbols above spectrum) determined from
the HITRAN database (Rothman et al. 1992). The diamonds indicate the
expected position of \eighteenco\ absorption lines, with the R(4) and
R(5) line detections labeled. The prominent broad absorption between
2123--2149 \waven\ can be fully ascribed to solid CO.  Wavelength
regions with poor atmospheric transmission are not
plotted.}~\label{f:obs}
\end{figure*}

The most critical step in the reduction of this data is correction for
atmospheric absorption features, and separate telluric and
interstellar CO absorption features. For our observations of L1489 IRS
this proved to be relatively easy, because at the time of our
observations the deep interstellar CO lines are shifted by as much as
43 \kms\ with respect to telluric CO lines.  The low airmass of the
source, 1.00--1.05, also made this step easier.  The standard stars HR
1380 (A7V) and HR 1497 (B3V) were observed at similar airmass, and
their spectral shape and hydrogen absorption features were divided out
with Kurucz model atmospheres.  A good telluric correction was
achieved, although residuals are seen near deep atmospheric lines. To
be safe, we therefore removed the parts of the spectrum which have
less than 50\% of the maximum transmission in each setting.  This does
not affect most of the features in the L1489 IRS spectrum, because the
velocity shift of 43 \kms\ {\it fully} separates them from telluric
features.  The final signal-to-noise value on the unsmoothed spectrum
is $\sim 45$, after integration times of 11, 14, and 20 minutes on
each of the three settings of this $M=4.8$ magnitude \citep{keny93}
source. The different settings were wavelength calibrated on the
atmospheric CO emission lines, and subsequently the three settings
were combined by applying relative multiplication factors.  We have
not attempted to flux calibrate the spectrum, since we are interested
in absorption features only.

\section{Results}

The fully reduced echelle spectrum of L1489 IRS shows, in great
detail, many deep narrow absorption lines of gas phase \twelveco\ and
\thirteenco, and a few weak lines of \eighteenco\
(Fig.~\ref{f:obs}). These lines were identified, using the line
frequencies in the HITRAN catalogue \citep{roth92}. The broad
absorption feature between 2122-2149 \waven\ can be attributed to the
stretching vibration mode of \twelveco\ in ices along this line of
sight.

In order to analyze these gas and solid state absorption features, a
shallow, second order polynomial continuum was applied to derive the
optical depth spectrum. The solid CO band was then analyzed using
available laboratory experiments (\S 3.2). The derivation of physical
parameters from the gas phase lines is highly model dependent.  First,
we will derive temperatures and column densities using the standard
curve of growth and rotation diagram techniques (\S 3.1). Then, we
will independently test an astrophysically relevant power law model in
\S 4.1. This information is combined in \S 4.3 to discuss the origin
and thermal history of the solid CO seen in this line of sight.

\subsection{Gas Phase CO}

The \twelveco\ lines have a complicated profile.  Deep lines are
present at a velocity of $+43$ \kms\ with respect to earth, which is
the systemic velocity of L1489 IRS at the date these observations were
done (taking $v_{\rm lsr}=$ 5 \kms\ from millimeter emission lines;
e.g. \citealt{hoge00a}). Each of the \twelveco\ lines is accompanied
by a fairly prominent wing at the red shifted side
(Fig.~\ref{f:comp1213}).  As an exploring step in the analysis, we
decomposed the main \twelveco\ component and its wing by fitting two
Gaussians. They are separated by on average 23$\pm$6 \kms, and the
main feature and its wing have widths of FWHM=20$\pm$3 \kms\ and
53$\pm$17 \kms\ respectively. The main \twelveco\ feature at the
systemic velocity is resolved.  Excess absorption is visible at the
blue and red sides with respect to \thirteenco\
(Fig.~\ref{f:comp1213}). The \thirteenco\ lines have Gaussian shapes
with a width equal to the instrumental resolution (FWHM=12 \kms).  The
main absorption feature of the \twelveco\ lines is therefore only in
part responsible for the same gas seen in \thirteenco.

\vbox{
\begin{center}
\leavevmode 
\hbox{%
\epsfxsize\hsize
\includegraphics[angle=90, scale=0.53]{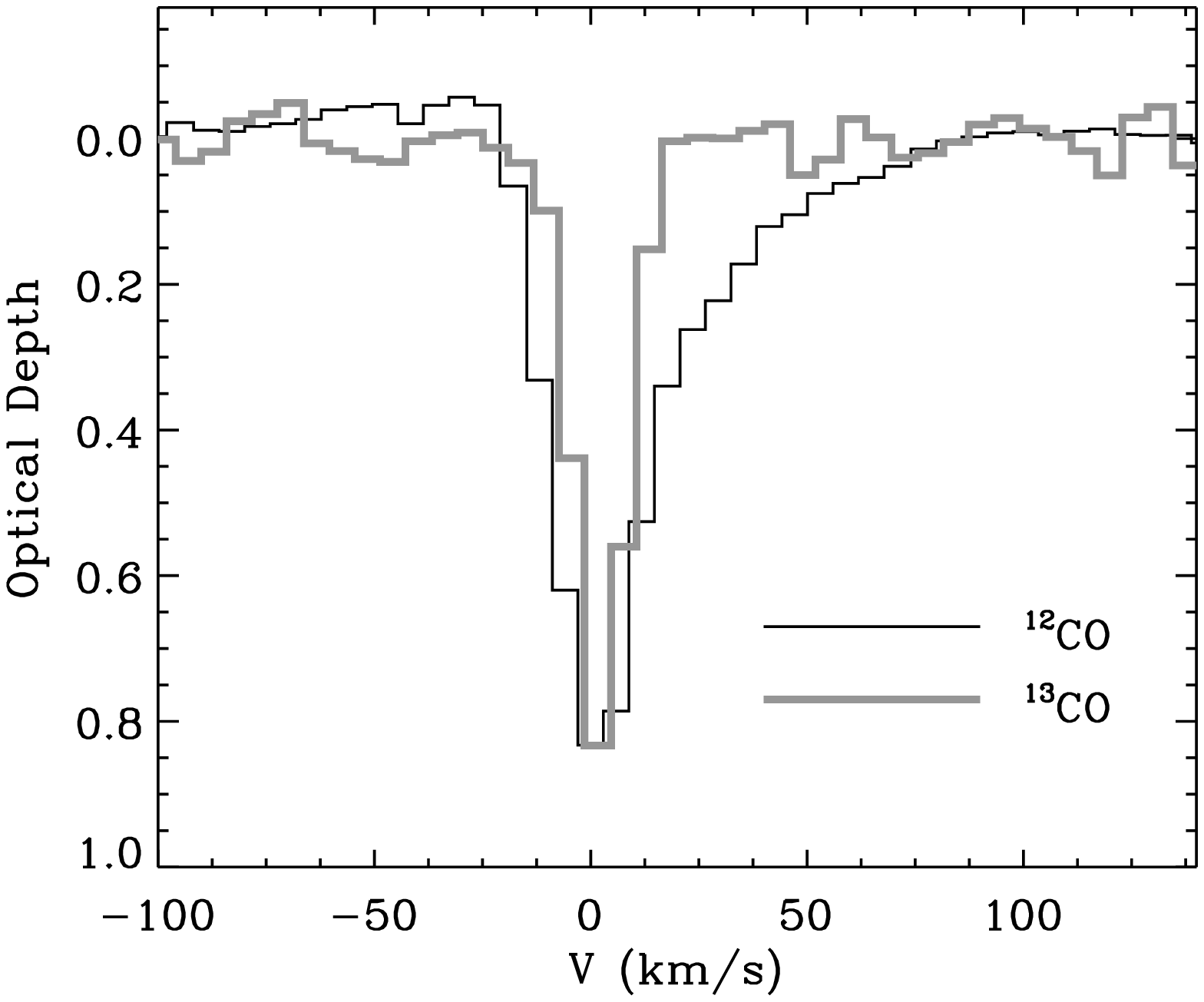}}
\figcaption{\footnotesize Comparison of the \twelveco\ and \thirteenco\ (thick gray
line) spectra of L1489 IRS. The spectra shown are averaged over the
observed \twelveco\ P(6)--P(15) lines and all \thirteenco\ lines in
order to increase the signal-to-noise. They are corrected for the
source and earth velocity (43 \kms) and the \thirteenco\ spectrum was
multiplied by 3.64 to facilitate comparison.~\label{f:comp1213}}
\end{center}}


We derived equivalent widths for the isotopes and \twelveco\
components (Table~\ref{t:eqw}), and applied a standard curve of growth
technique to calculate column densities for each $J$ level.  The
comparison of equal $J$ levels of \eighteenco, \thirteenco, and main
\twelveco\ component provides a handle on the intrinsic line width
\bdop\ (=FWHM/$2\sqrt{\rm ln 2}$), in the assumption that all material
absorbs at the same velocity (however, see \S 4.1). One also has to
assume that the isotope ratios are constant along the line of sight
(\twelveco/\thirteenco=80 and \twelveco/\eighteenco=560;
\citealt{wils94}). The equivalent widths of \eighteenco\ and
\thirteenco\ are then simultaneously fit at \bdop$>$0.8 \kms.  Lower
\bdop\ significantly ($>3 \sigma$) underestimates the \thirteenco\
lines, with respect to \eighteenco. As mentioned above, the main
\twelveco\ component is clearly contaminated by gas not seen in
\thirteenco, both on the blue and red shifted sides.  As a first order
correction we lowered the \twelveco\ equivalent widths in
Table~\ref{t:eqw} with 40\%, which corresponds to unresolved lines at
the observed peak optical depth.  Then, we find that all isotopes are
best fit simultaneously at \bdop=1.4$\pm 0.1$ \kms. This would be an
upper limit if the correction for contamination of the \twelveco\
lines were too small.  Hence, this curve of growth analysis of the CO
isotopes shows that, in the assumption that all gas absorbs at the
same velocity, \bdop\ is limited to 0.8$<$ \bdop $<$1.5 \kms.

\begin{table*}[t!]
\caption{Equivalent Widths}~\label{t:eqw}
\footnotesize
\center
\begin{tabular}{ccccccc}
\tableline
\noalign{\smallskip} 
transition       &  \multicolumn{2}{c}{\eqw}               & transition  &  \eqw              & transition       &  \eqw\\
\twelveco        &   \multicolumn{2}{c}{10$^{-3}$ \waven}  & \thirteenco &  10$^{-3}$ \waven  & \eighteenco      &  10$^{-3}$ \waven  \\
                 & main  & wing                            &             &                    &                  &      \\
\noalign{\smallskip}  
\tableline
\noalign{\smallskip} 
R(5)    &88 (11)&70 (22)& R(17) & $<$ 3         & R(6)  & $<$3          \\
R(4)    &84 (10)&54 (17)& R(16) & $<$ 5         & R(5)  & 6 (2)         \\   
R(3)    &67 (9) &55 (18)& R(15) & 5 (2)         & R(4)  & 6 (2)         \\
R(2)    &73 (10)&64 (20)& R(14) & \nodata       & R(3)  &$<5$           \\
R(1)    &70 (9) &39 (12)& R(13) & 5 (2)         & R(2)  & \nodata       \\
R(0)    &68 (9) &\nodata & R(12)& $<$ 4         & R(1)  & \nodata       \\
P(1)    &65 (9) &11 (4) & R(11) & 6    (2)      & R(0)  & $<6$          \\
P(2)    &72 (10)&30 (10)& R(10) & 6    (2)      & P(1)  & $<3$          \\
P(3)    &80 (11)&21 (7) & R(9)  & 7    (2)      & P(2)  & $<3$          \\
P(4)    &69 (9) &82 (26)& R(8)  & \nodata       & \nodata & \nodata     \\
P(5)    &84 (11)&73 (23)& R(7)  & 12   (3)      & \nodata & \nodata     \\
P(6)    &85 (10)&65 (20)& R(6)  & 13   (2)      & \nodata & \nodata     \\
P(7)    &85 (11)&90 (28)& R(5)  & 19   (2)      & \nodata & \nodata     \\
P(8)    &85 (11)&71 (22)& R(4)  & 17   (2)      & \nodata & \nodata     \\
P(9)    &85 (11)&69 (21)& R(3)  & 21   (2)      & \nodata & \nodata     \\
P(10)   &86 (10)&84 (26)& R(2)  & 18   (2)      & \nodata & \nodata     \\
P(11)   &79 (10)&61 (18)& R(1)  & \nodata       & \nodata & \nodata     \\
P(12)   &79 (11)&56 (17)& R(0)  & 24   (2)      & \nodata & \nodata     \\
P(13)   &88 (12)&43 (13)& P(1)  & 22 (2)        & \nodata & \nodata     \\
P(14)   &69 (9) &64 (19)& P(2)  & 25 (2)        & \nodata & \nodata     \\
P(15)   &72 (10)&53 (16)& P(3)  & 18 (2)        & \nodata & \nodata     \\
P(16)   &\nodata &\nodata & P(4)& 15 (2)        & \nodata & \nodata     \\
\noalign{\smallskip} 
\tableline
\end{tabular}
\end{table*}

\begin{table*}
\caption{Physical Parameters From Curve of Growth and Rotation Diagram}~\label{t:phys}
\center
\footnotesize
\begin{tabular}{lcccl}
\tableline 
\noalign{\smallskip} 
\trot & \bdop & $N$(\twelveco)$^a$ & $v_{\rm lsr}$ &Origin\\ 
\noalign{\smallskip} 
K & \kms & $10^{18}$ \sqcm & \kms & \\ 
\noalign{\smallskip} 
\tableline 
\noalign{\smallskip}
19$^{+5}_{-4}$ & 1.6 & 6 & 5$\pm$3 & circumstellar disk+foreground; from \thirteenco \\ 
15$^{+8}_{-3}$ & 1.3 & 10 & 5$\pm$3 & as above, but alternative \bdop \\ 
13$^{+8}_{-3}$ & 1.0 & 19 & 5$\pm$3 & as above, but alternative \bdop \\ 
11$^{+8}_{-3}$ & 0.7 & 96 & 5$\pm$3 & as above, but alternative \bdop \\
 & & & & \\ 
300$^{+300}_{-150}$ & 1.3 & 2.4& 5$\pm$3 & disk; from \thirteenco\\ 
250$^{+300}_{-100}$ & 1.0/0.7 & 2.8 & 5$\pm$3 & as above, but alternative \bdop \\
 & & & & \\ 
250$^{+500}_{-100}$ & $<32$ & 0.20$\pm 0.03^b$ & 28$\pm$6 & \twelveco\ red wing\\
 & & & & \\
500$\pm 250$ & $<12$ & $\sim 0.06^b$ & $-$5/15 &  blue and red wings resolved \twelveco\ main\\
\tableline
\noalign{\smallskip} 
\multicolumn{5}{l}{$^a$ assuming N(\twelveco)/N(\thirteenco)=80}\\ 
\multicolumn{5}{l}{$^b$ assuming optically thin absorption}\\
\end{tabular}
\end{table*}

With the column densities per $J$ level at hand, a rotation diagram
was constructed for the \thirteenco\ lines (Fig.~\ref{f:rot}) to
derive the total column density and temperature at a number of allowed
\bdop\ values (Table~\ref{t:phys}). Clearly, the rotation diagram
shows a double temperature structure, much resembling that of high
mass objects \citep{mitc90}: cold ($T\sim 15$ K), and warm gas ($T\sim
250$ K) are present along the same line of sight.  The column density
of the cold component toward L1489 IRS is a particularly strong
function of \bdop, increasing with an order of magnitude from 0.7 to
1.3 \kms. An independent CO column of 1.4$\times 10^{19}$ \sqcm\ can
be estimated from $A_{\rm V}$=29 \citep{myer87}, assuming the dust and
gas are co-spatial. This would suggest \bdop\ is in the 1.0-1.3 \kms\
range.

\vbox{
\begin{center}
\leavevmode 
\hbox{%
\epsfxsize\hsize 
\includegraphics[angle=90, scale=0.49]{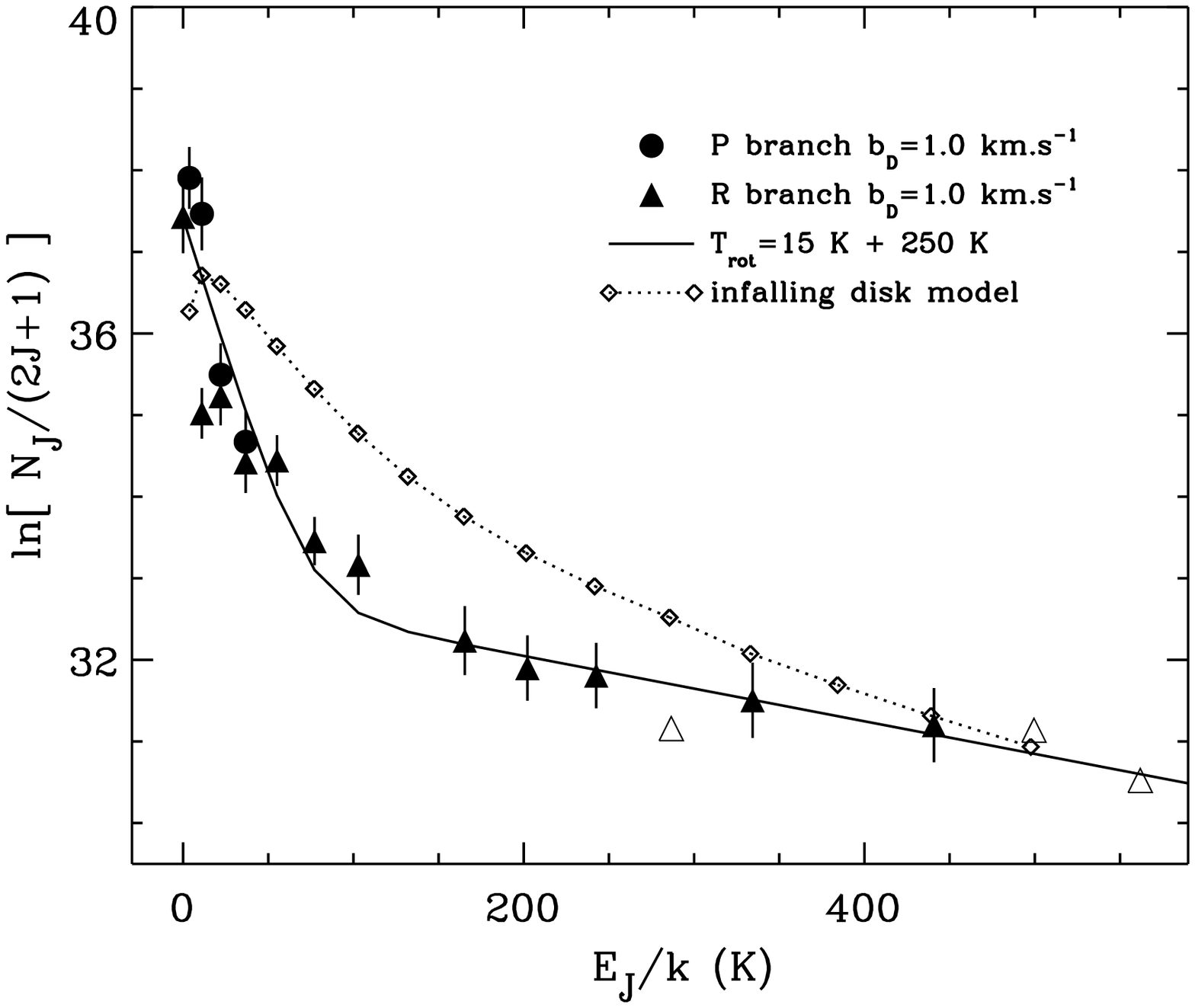}} 
\figcaption{\footnotesize Rotation diagram of \thirteenco, with the
column densities on the vertical axis calculated from the curve of
growth with \bdop=1.0 \kms\ (dots: P branch; closed triangles: R
branch; open triangles: R branch upper limits).  The solid line
corresponds to temperatures $T_{\rm rot}$=15 K, and 250 K
(Table~\ref{t:phys}) in the Boltzmann equation. Open diamonds
connected with a dotted line represent the rotation diagram for our
disk model, which is different from the curve of growth analysis
because of a lower assumed \bdop\ (0.1 \kms) and the presence of a
velocity gradient (\S 4.1), although both models fit the observed
\thirteenco\ lines well.~\label{f:rot}}
\end{center}}


For the highly red shifted \twelveco\ wing a rotation diagram is
constructed as well. Here, we assume that the absorption is optically
thin, because the wings are absent in \thirteenco, as evidenced by the
high signal-to-noise average line profile (Fig.~\ref{f:comp1213}).
The rotational temperature of this highly red shifted gas is similar
to that of the warm \thirteenco\ component (250$^{+500}_{-100}$ K),
however the column is an order of magnitude lower
(Table~\ref{t:phys}). Finally, assuming that the absorption in the
wings of the resolved \twelveco\ main component is optically thin, the
column of this blue and red shifted low velocity gas (within 10 \kms\
of the systemic velocity) is about 25\% of the column of the high
velocity red shifted gas at a similar, although poorly determined,
temperature.

Summarizing, this basic analysis indicates that, from \thirteenco,
both a large column of cold 15 K gas and a significant amount of warm
gas ($T\sim250$ K) are present within $\sim 3$ \kms\ of the systemic
velocity (Table~\ref{t:phys}). The \twelveco\ lines show that warm gas
at $T\sim250$ K is also present at {\it highly red shifted} velocities
($20-100$ \kms), but at a factor 10 lower column. A small amount of
warm gas is present at {\it low red and blue shifted} velocities as
well (within 10 \kms). As further described in \S 4.1, the gas
components at the red shifted and systemic velocities can be fitted
within the same physical model of a contracting disk, but the origin
of the warm gas at low blue shifted velocities is more difficult to
explain.

\subsection{Solid CO}

The broad absorption feature between 2122-2149 \waven\
(Fig.~\ref{f:obs}) can be entirely attributed to the stretching
vibration mode of \twelveco\ in circumstellar ices (\S 4.3). The high
spectral resolution allows, for the first time, to unambiguously
separate the gas phase CO lines from the solid state absorption and
study the solid CO band profile in great detail. In accordance with
previous, low resolution studies \citep{chia98, teix98}, a distinct
narrow feature is seen at 2140 \waven, and a significantly broader
component at longer wavelengths. Our data however, indicates the
presence of a new, third component on the blue side, separate from the
narrow 2140 \waven\ feature, most notable by a change of the blue
slope at 2142 \waven\ (Fig.~\ref{f:obs}).

In order to explain the shape of this CO absorption profile, we have
taken laboratory experiments of solid CO from the literature
(\citealt{sand88, schm89, palu93, trot96, ehre97}; Elsila,
Allamandola, \& Sandford 1997; \citealt{bara98, teix98}). We
determined peak position and width of the laboratory profiles as a
function of ice composition, temperature, and cosmic ray bombardment
intensity.  Additionally, for CO--rich ices we used optical constants
to calculate the absorption profile as a function of particle shape in
the small particle, Rayleigh limit (for details, see
\citealt{ehre97}). The absorption profile of CO--poor ices
(concentration $<$30\%) is not affected by these particle shape
effects.  Changing these various parameters gives a wide variety of
peak positions (2135--2144 \waven) and widths (FWHM=1.5--14 \waven) in
the laboratory.  We limit ourselves here by identifying general
trends, as summarized in Fig.~\ref{f:lab}. These trends are then
related to the Gaussian peak position and width of the three
aforementioned components observed toward L1489 IRS (also given in
Fig.~\ref{f:lab}).

\vbox{
\begin{center}
\leavevmode 
\hbox{%
\epsfxsize\hsize 
\includegraphics[angle=90, scale=0.49]{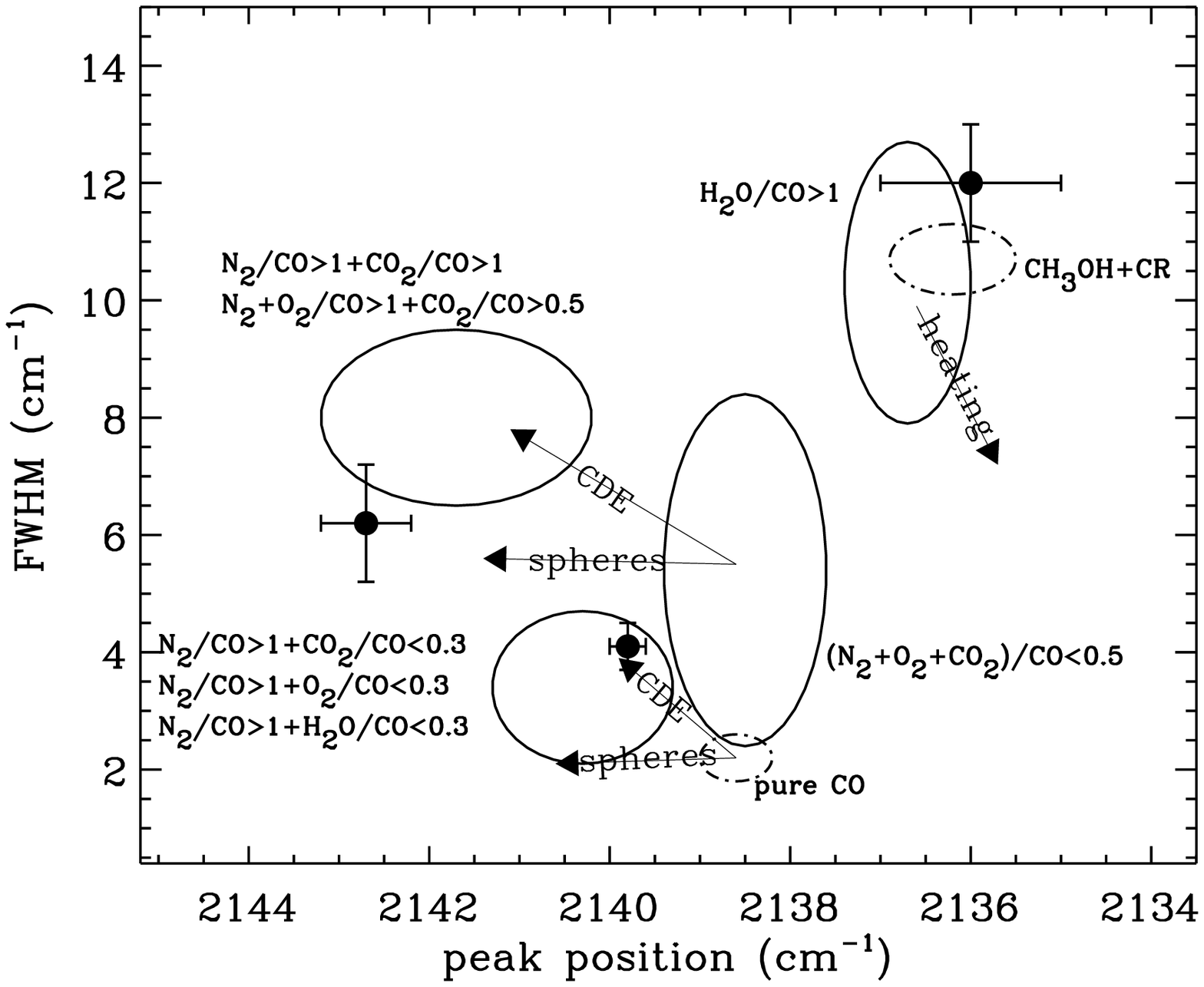}} 
\figcaption{\footnotesize Diagram showing schematically the effect of several
astrophysically relevant parameters on the peak position and width of
the solid $\rm ^{12}CO$ stretching mode, as determined from laboratory
experiments.  Ellipses indicate ices of various composition.  For
CO--rich ices, the effect of grain shapes is indicated by arrows for
spheres and ellipsoids (`CDE').  For polar ices, the effect (magnitude
and direction) of heating is given by the arrow.  `CH$_3$OH+CR' means
a CH$_3$OH or a CH$_3$OH:H$_2$O ice irradiated by energetic particles
simulating cosmic rays.  The dots with error bars represent the
observed peak position and width of the absorption components detected
toward L1489 IRS.~\label{f:lab}}
\end{center}}


The absorption band of a thin film of pure, solid CO at $T=$10 K peaks
at $\sim$2137 \waven, and is very narrow ($\sim$2 \waven).  This
clearly does not fit any of the components observed toward L1489 IRS.
However, for ellipsoidally shaped grains (more precisely, a
distribution of ellipsoidal shapes, `CDE'; \citealt{bohr83}), this
sensitive strong band shifts to shorter wavelengths and becomes
broader. Now it exactly fits the 2140 \waven component observed toward
L1489 IRS, both in peak position and width (Fig.~\ref{f:lab}). In view
of a recent controversy on optical constants \citep{ehre97}, it is
worth noting that good fits are obtained with optical constants from
the works of \citet{ehre97} and \citet{bara98}. The optical constants
of \citet{trot96} do not induce strong particle shape effects and
therefore pure CO does not provide a good fit to L1489 IRS for any
particle shape.
 
Broadening of the laboratory profile, in order to fit the 2140 \waven\
feature in L1489 IRS, is also achieved by adding a small amount of
CO$_2$, O$_2$, or H$_2$O molecules. To avoid a too large broadening,
and minimize aforementioned particle shape effects, this mixture needs
be diluted in N$_2$. This astrophysically relevant molecule does not
broaden the feature, and gives a small blue shift \citep{ehre97,
elsi97}, required to fit the 2140 \waven\ feature in L1489 IRS. Thus,
both this mixture, as well as ellipsoidally shaped pure CO ice grains,
provide good fits to the central 2140 \waven\ feature.

Now, the width significantly increases, and the peak shifts to longer
wavelengths by diluting CO in a mixture of molecules with large dipole
moments such as H$_2$O or CH$_3$OH \citep{sand88, tiel91}.  This
particular behavior is needed to fit the broad long wavelength wing
seen toward L1489 IRS. Solid CH$_3$OH has an abundance less than a few
percent of solid H$_2$O toward low mass objects \citep{chia96}. H$_2$O
seems the best dilutant, because of its large interstellar abundance.
The use of H$_2$O:CO mixtures requires interesting additional
constraints.  Although a low temperature, unprocessed H$_2$O--rich ice
does provide a good fit to the red wing, it can be excluded based on
the presence of a prominent second absorption at $\sim$2150 \waven\ in
the laboratory, which is clearly absent toward L1489 IRS.  This second
peak is caused by CO molecules located in pockets in an amorphous
ice. These CO molecules are weakly bound, and the $\sim$2150 \waven\
peak disappears rapidly at higher $T$ or as a result of cosmic ray
hits \citep{sand88}. Thus, the H$_2$O ice responsible for the long
wavelength wing toward L1489 IRS must be thermally ($T>$ 50 K) or
energetically processed.

The blue wing seen at $\sim$2143 \waven\ toward L1489 IRS can only be
explained by an apolar ice. Adding a significant amount of CO$_2$ to a
CO ice (CO$_2$/CO$>1$) results in the blue shift and broadening
required to fit the observed wing. Somewhat less CO$_2$ is needed
(CO$_2$/CO$\sim 0.5$) when a large amount of O$_2$ is present.  N$_2$
may be added as well, but is not essential except as a dilutant to
reduce the effects of particle shape.  A good fit is obtained by the
mixture N$_2$:O$_2$:CO$_2$:CO=1:5:0.5:1, as proposed in
\citet{elsi97}.  If CO$_2$/CO$<0.5$ the band peaks at too high
wavelength. In CO--rich ices this problem can however be overcome by
particle shape effects (Fig.~\ref{f:lab}). In view of this effect it
is not possible to constrain the relative molecular abundances of this
interstellar component in more detail, but it is clear that an apolar
CO$_2$ or O$_2$ ice is needed, different from the distinct 2140
\waven\ feature.

A three component fit to the entire CO ice band of L1489 IRS is shown
in Fig.~\ref{f:coice}. Although this is not a unique fit, it does obey
the global trends that we identified in the laboratory experiments.

Finally, the solid CO column density is derived by dividing the
integrated optical depth over the band strength $A$. We take $A =
1.1\times 10^{-17}$ cm molecule$^{-1}$ independent of ice composition
\citep{gera95}, and thus find $N$(solid CO)=6.5$\times 10^{17}$
\sqcm. The main source of uncertainty here is in $A$, which is about
10\%. CO in polar ices contributes 3.5$\times 10^{17}$ \sqcm\ to the
total column and the apolar components at 2140 and 2142 \waven\
contribute each 1.5$\times 10^{17}$ \sqcm.

\vbox{
\begin{center}
\leavevmode 
\hbox{%
\epsfxsize\hsize 
\includegraphics[angle=90, scale=0.49]{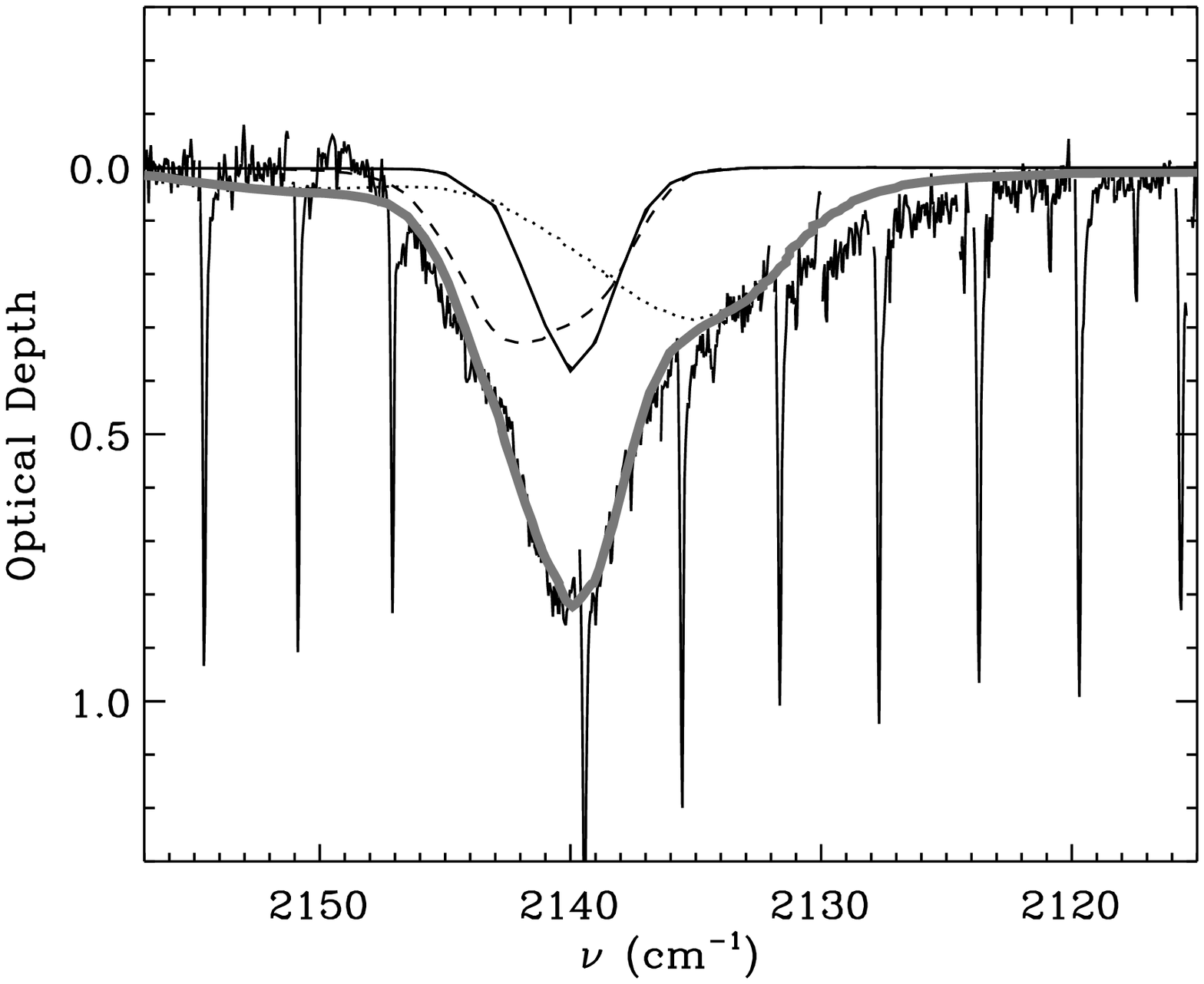}} 
\figcaption{\footnotesize Observed CO absorption band of L1489 IRS, fitted by a
combination of three laboratory ices: a polar H$_2$O:CO=4:1 ($T$=50 K;
dotted) to account for the long wavelength wing, an apolar
N$_2$:O$_2$:CO$_2$:CO=1:5:0.5:1 ($T$=10 K; dashed) for the short
wavelength wing, and a pure CO ice ('CDE' shape; $T$=10 K; solid) that
fits the central peak.  The thick, smooth gray line is the sum of
these components.~\label{f:coice}}
\end{center}}


\section{Discussion}

\subsection{An Infalling Disk}

The astrophysical meaning of the apparent two component temperature
structure seen in the \thirteenco\ rotation diagram (Fig.~\ref{f:rot})
requires further investigation. For high mass protostars it was found
that similar rotation diagrams can be `mimicked' by power law models
of spherical envelopes \citep{tak00}.

For L1489 IRS, the detection of molecular gas at a range of
temperatures and red shifted velocities could indicate the presence of
infalling gas at a range of radii from the protostar.  Indeed, a
2000~AU radius contracting, disk-like structure was found in
millimeter wave interferometer data \citep{hoge00a}. In a detailed
follow-up study, \citet{hoge01} adopts a flared-disk model based on
\citet{chia97} with a radial power-law distribution for the
temperature
\begin{equation}
T = 34(R/1000\,{\rm AU})^{-0.4} {\rm ~K,}
\end{equation}
and a density distribution that has a power-law drop-off with radius
and a vertical exponential drop-off with scale height $h$
\begin{equation}
\rho(R,z) =\rho_0 (R/1000 {\rm AU})^{-1.5} {\rm e}^{-z^2/h^2} {\rm ~kg~cm^{-3}}.
\label{eq:den}
\end{equation}
The scale-height $h$ is assumed to be a simple function of R, $h=R/2$.
An inward-directed radial velocity field described as
\begin{equation}
V_{\rm in} = 1.3 (R/100\,{\rm AU})^{-0.5} {\rm ~km~s^{-1}}
\end{equation}
is inferred, in addition to Keplerian rotation around a 0.65~M$_\odot$
central star.

\begin{figure*}[t!]
\center
\includegraphics[angle=90, scale=0.72]{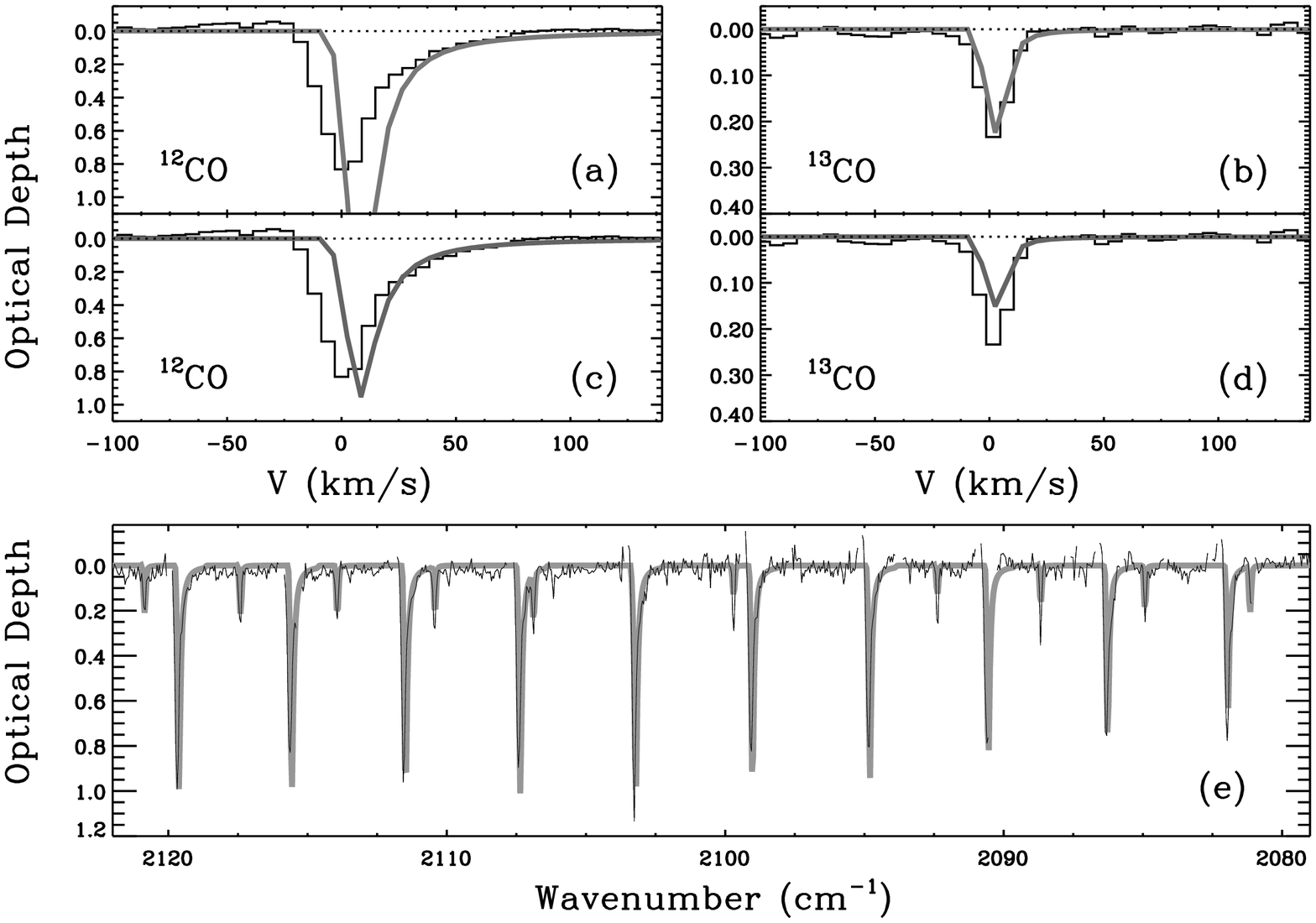}
\caption{Comparison of the observed spectrum of L1489 IRS with an
infalling disk model. The histograms in panels (a) and (b) represent
the average of the observed \twelveco\ P(6)--P(15) lines and all
\thirteenco\ lines respectively, corrected for the source and earth
velocity (43 \kms).  The smooth thick gray line in these two panels is
the collapsing disk model, averaged over the same lines.  In panels
(c) and (d) the same averaged observed spectra and model are plotted,
but now assuming that in the model 1\% of the unextincted stellar
continuum emission does not pass through the disk.  Panel (e) shows
the latter model (thick gray line) compared to a portion of the
non-averaged L1489 IRS spectrum.}~\label{f:avemod}
\end{figure*}

Can this contracting disk model, based on (sub-) millimeter emission
observations with angular resolution of $4$--$8''$, reproduce the
observed infrared CO absorption line profiles measured along a pencil
beam?  The absorption lines are modeled with the radiative transfer
code of \citet{hoge00b}; the high densities in the disk ensure LTE
excitation for the lines involved, and line trapping is neglected in
the excitation calculation.  The model spectra include dust opacity at
a standard gas/dust ratio, as well as a $N({\rm CO})=1\times 10^{18}$
cm$^{-2}$ column of cold foreground material (15 K; \citealt{hoge01});
both factors do not affect the spectra in any significant way.  The
calculated spectrum is convolved with a Gaussian of FWHM=12 \kms,
which is the NIRSPEC instrumental resolution.

We find that, while keeping all other parameters the same as in
\citet{hoge01}, the assumed density profile sensitively influences the
wings of the $^{12}$CO lines. This is enhanced by the fact that we are
observing the flared disk of L1489 IRS at an inclination between
$60^\circ$ and $<90^\circ$ (cf., \citealt{padg99}), and the pencil
beam crosses the disk at a few scale heights.  Small changes in the
density profile, for example induced by the thermal structure, have a
large effect on the absorption line profile.  In the model of
\citet{hoge01} the scale height increases linearly with distance from
the star, and thus the density $\rho (l)$ along the line of sight $l$
follows the density in the mid-plane (Eq.~\ref{eq:den}) reduced by a
factor $e^{-4/\tan^2(\alpha)}$, with $\alpha$ the inclination. Here,
we include the effect of density variations, or deviations from the
adopted scale height $h=R/2$, by relaxing the values of the density
along the line of sight $l$ by fitting $\rho(l) = \rho_0 (l/1000 {\rm
~AU})^{-p}$ to the data.

This initial model successfully fits the peak velocity and depth of
both high and low $J$ $^{13}$CO lines (Fig.~\ref{f:avemod}).  Its
rotation diagram is quite different from that of the curve of growth
analysis (Fig.~\ref{f:rot}), showing that rotation diagrams must be
interpreted with great care.  Our model also matches the range of
velocities observed in the red wings of the $^{12}$CO lines, when
taking $p=0.55\pm 0.15$. This is a much shallower density profile
compared to that derived from millimeter wave data ($p=1.5$;
Eq.~\ref{eq:den}) and indicates that the scale height increases more
than linear, i.e. the disk flares more than assumed in \citet{hoge01}.
With this result, it is possible to determine the important relation
of disk scale height $h(R)=a.R^b$ as a function of $R$, but only if
the disk inclination is {\it a priori} known. Unfortunately the
inclination is not better constrained than within the range of
$60^\circ$ and $<90^\circ$ imposed by near-infrared data
(\citealt{padg99}). We can therefore not distinguish between low and
high values of $a$ and corresponding high and low inclinations
respectively. In either case, the total $^{12}$CO column along the
pencil beam is $1.2\times 10^{19}$ \sqcm, with 58\% of the CO mass at
a temperature of $T=20-60$ K, 15\% at $60-90$ K, and 27\% at $60-90$
K. This result is of importance in \S 4.3 in the interpretation of the
solid CO observations, and in particular in assessing the thermal
history of ices.  The total column of our model is in good agreement
with the column derived from the visual extinction ($1.4\times
10^{19}$ \sqcm; \S 3.1).  It is also of the same order of magnitude as
the total column through the mid-plane, calculated from dust and line
emission ($N[{\rm CO}]=6\times 10^{18}$ \sqcm; \citealt{hoge01}), and
confirms the relatively edge-on orientation of the disk.

However, apart from these successes, the \twelveco\ lines show that
our infalling disk model produces too much warm gas at high velocities
(Fig.~\ref{f:avemod}). The $^{12}$CO lines are a factor of 2.5 deeper,
and, in contrast to $^{13}$CO, they peak at a too high velocity (+10
\kms) with respect to the observations.  In principle, one could make
the \twelveco\ lines less deep by assuming that $\sim 1\%$ of the
original, unextincted continuum flux (corresponding to 30\% of the
extincted continuum) reaches the slit without passing through the
disk, by scattering on large grains.  The shift in peak velocity
however requires a solution of a more fundamental origin.  Perhaps the
infall velocity function is shallower, and the disk is more
rotationally supported at lower radii.  The amount of warm gas at high
velocities can also be lowered by assuming that only part of the disk
participates in the high velocity inflow, such as a thin hot surface
layer, or gas accelerated in magnetic field tubes directed from the
inner disk to the stellar photosphere.  Such a two component model is
consistent with the rotation diagram derived from the curve of growth
(Fig.~\ref{f:rot}), and also with the low observed mass accretion
rate.  If we take the inflow at face value, and assume that the entire
disk participates, the mass accretion rate would be $10^{-6}$
M$_\odot$, generating 7 L$_\odot$ in accretion luminosity. The star's
L$_{\rm bol}$ is estimated at 3.7 L$_\odot$ which also contains the
stellar luminosity. It is therefore indeed likely that the mass
accretion onto the star is significantly lower, as is also traced
through the lack of the hydrogen Pf$\beta$ emission line in our
spectrum (2148.8 \waven; Fig.~\ref{f:obs}) and the weakness of
Br$\gamma$ emission \citep{muze98}.

\subsection{Binarity?}

An entirely different explanation for the line profiles may lie in the
possibility that L1489 IRS is a protobinary system.  A protobinary
nature of L1489 IRS is suggested by various pieces of evidence
(\citealt{luca00}; \citealt{wood01} and references therein).  The
presence of a quadrupolar outflow system is inferred from K band
polarization images, C$^{18}$O emission line profiles, Herbig-Haro
knots that are scattered throughout the L1489 IRS environment, and a
very complex near infrared scattered light pattern.  Three dimensional
models, in which the axisymmetry of the infalling circumstellar
envelope is broken by multiple outflow cavities that are perpendicular
to each other, are able to account for the observed morphology.

The putative binary itself, however, has not been resolved so far.  An
upper limit on the projected separation from near infrared images has
been set on $<$ 20 AU \citep{padg99}.  If the CO absorption line
profile is in any way related to a binary system, then the large
observed velocities ($\sim$23 \kms; \S 3.1) may indeed favor a close
binary system. The line profile is then expected to vary on a time
scale of a few months, which can easily be tested.  In this case, much
of the observed warm gas might be present in two small circumstellar
disks, which are in a close orbit around each other. Some of the warm
gas may also be present at low density in the central cavity created
by the binary.  The large column of cold gas may originate in the
circumbinary disk. The binary extracts momentum from the 2000 AU
circumbinary disk, setting up the inward motion seen in millimeter
wave emission lines.  We leave further investigation of this topic for
future studies.

\subsection{The Origin and Evolution of Ices}

In order to establish if the solid CO observed toward L1489 IRS
originates in foreground clouds or in a circumstellar (or binary)
disk, it is worth to compare with ices observed in lines of sight not
affected by star formation. Observations of field stars obscured by
intervening quiescent material of the Taurus Molecular Cloud have
revealed that solid CO is not present when the extinction $A_{\rm
V}\lesssim 5$ (e.g. \citealt{teix99}). The solid CO toward L1489 IRS
can therefore not be associated with foreground clouds, which have a
gas column of $N$($^{12}$CO)= 1$\times 10^{18}$ \sqcm\ \citep{hoge01},
corresponding to $A_{\rm V}\sim 2$.

Thus, the solid CO must be present in the disk of L1489 IRS.  The
absorption profile is intriguingly different from that seen in
quiescent clouds.  The broad red wing has a depth of $\sim$30\% with
respect to the narrow 2140 \waven\ peak, which is significantly more
than toward all measured background stars (10\%; \citealt{chia95}).
This may well be an effect of thermal processing along the L1489 IRS
line of sight, because the sublimation temperatures of polar and
apolar ices, causing the broad and narrow features respectively, are
very different (90 versus 18 K).  However, a chemical origin of an
increased abundance of polar ices in disks cannot be excluded, because
the apparently edge-on system Elias 18 in the Taurus Molecular Cloud
has an extremely large CO depletion factor
(solid/[gas+solid]$\sim$100\% versus 7\% for L1489 IRS), but a deep
red `polar' CO wing is present as well \citep{shup01, chia98}. On the
other hand, energetic processing may take place even in the cold disk
of Elias 18 \citep{whit01}.  Clearly, it is necessary to
observationally characterize the ices in circumstellar disks in much
more detail.

If for now we assume the sublimation scenario, we can do some general
extrapolations which can be compared with the results of our gas phase
study (\S 4.1).  By scaling the long wavelength wing of solid CO of
background field stars to that of L1489 IRS, we find that a column of
6$\times 10^{17}$ \sqcm\ of CO has evaporated from the apolar ice
component in the part of the L1489 IRS disk along the pencil
absorption beam where $T<90$ K (the sublimation temperature of polar
ices).  Then the column of solid CO that went from the quiescent cloud
into building this part of the disk is 12.5$\times 10^{17}$ \sqcm.
Extrapolating this further, we use the observed CO depletion factor of
30\% toward field stars behind the Taurus Molecular Cloud
\citep{chia95} to calculate that the original quiescent gas column
must have been of the order of 3$\times 10^{18}$ \sqcm. Adding the
evaporated column, the expected present day gas column at $T< 90$ K is
3.6$\times 10^{18}$ \sqcm. This is of the same order of magnitude as
the CO column below 90 K in our collapsing disk model
($N$[CO]=8.7$\times 10^{18}$ \sqcm), which may indicate that no
chemical change in the apolar/polar CO ice ratio and no significant
additional depletion has occurred in the evolution from quiescent
Taurus Molecular Cloud material to the formation of the L1489 IRS
disk. This contrasts strongly with the very large depletions found in
the (older) disks of T Tauri stars \citep{dutr97}. The low CO
depletion along the pencil beam toward L1489 IRS (7\%) and the
supposed signs of thermal processing (see below) may be due to the
fact that our line of sight does not cross the disk mid-plane,
i.e. the system is not exactly edge-on.  The ice processing we see
takes place higher in the disk atmosphere, perhaps in the warm layer
below the super-heated dust layer responsible for millimeter wave line
emission \citep{zade01}. It must be noted that in the model of
\citet{hoge01} the gas temperatures are larger than 25 K, prohibiting
the formation of apolar ices and large CO depletions anywhere in the
disk. The observed presence of apolar CO ices thus indicates that, as
already suggested in \S 4.1, the line of sight may cross the cold,
rotationally supported disk interior not traced in the observations and
infall model of \citet{hoge01}.

Apart from evaporation of apolar ices, other hints of thermal
processing include the aforementioned absence of the 2150 \waven\
absorption (\S 3.2), which occurs in cold unprocessed polar CO ices
but disappears at temperatures $T>50$ K. Also, the blue apolar wing
may be a consequence of thermal processing. If the central 2140
\waven\ peak is due to a mixture of O$_2$, N$_2$ and CO instead of
pure CO (spectroscopically these cannot be distinguished), thermal or
energetic processing (UV radiation from the ISRF, UV induced by H$_2$
cosmic ray collisions, or direct hits of cosmic rays) could
efficiently produce CO$_2$. This could cause the band to broaden and
shift to the position of the observed blue wing. Chemical models
indicate that energetic processing of molecules in disks takes place
on a time scale of 10$^6$ yrs \citep{aika99}, which is somewhat longer
than the age of the disk of L1489 IRS ($\sim 5\times 10^5$ yrs).  This
however applies to the disk mid-plane, and the time scale may well be
shorter in the lower density higher disk layers that our observations
of L1489 IRS trace.  A possible problem with the energetic processing
interpretation is the absence of a feature adjacent to the short
wavelength side of the CO ice band, usually attributed to
energetically produced C$\equiv$N bondings \citep{whit01}.  Another
spectroscopic tracer of thermal processing is the signature of
crystallization in the band profiles of H$_2$O and CO$_2$ ices. Our
infalling disk model predicts that only 15\% of the gas is within the
temperature range at which ices crystallize (60--90 K), and thus
crystallization is not expected to play a significant role in the disk
of L1489 IRS. This model prediction can be tested with future high
quality H$_2$O and CO$_2$ spectra of L1489 IRS.

In summary, several pieces of evidence indicate that the CO ices in
the disk of L1489 IRS have experienced thermal or energetic
processing.  The strongest arguments are the low depletion factor and
the low ratio of apolar to polar ices with respect to the quiescent
Taurus Molecular Cloud material. This may be explained by the fact
that the disk of L1489 IRS is seen under an angle, and our pencil
absorption beam traces the warm upper disk layers.

\vspace{15pt}
\section{Summary and Future Work}

We have shown that valuable and unique information is obtained from
high resolution spectroscopy of the CO fundamental at 4.7 \mum\ toward
the low mass class I protostar L1489 IRS in the Taurus Molecular
Cloud. At a resolution of $R=25,000$ (12 \kms) this object shows a
multitude of deep ro-vibrational absorption lines of \twelveco, as
well as \thirteenco\ and \eighteenco. The isotopes trace large columns
of warm and cold gas in the circumstellar disk at or within 3 \kms\ of
the systemic velocity, while the \twelveco\ line profiles show warm
gas that is red shifted at a range of velocities of up to 100 \kms.
Both the line depth of the isotopes and the extent of the red shifted
warm gas seen in \twelveco\ are well explained by an infalling flared
disk model with power laws for the temperature, infall velocity, and
density (a small column of blue shifted gas seen in the \twelveco\
line wing however remains unexplained).  These observations show that
the inward motions inferred on scales of several hundred AU through
millimeter wave interferometry, continue to within 0.1 AU of the star,
where the velocity model of \citet{hoge01} predicts inward velocities
exceeding several tens of \kms.  A detailed comparison of our power
law infall model however overestimates the amount of warm, high
velocity infalling gas.  Much of this gas must therefore be
rotationally supported, and only a thin disk surface layer is
infalling, or gas is accelerated along magnetic fields in the inner
parts of the disk. High spatial resolution millimeter wave
observations (with ALMA) are needed to test our model, e.g. to refine
the determination of the velocity field in the inner disk parts and
the dependence of the disk scale height on radius $h(R)$.  Finally,
high spatial resolution infrared interferometer observations would be
able to see if L1489 IRS is a close binary system ($<$ 20 AU), which
is essential to assess the importance of this aspect to the observed
line profiles and disk evolution.

In the same spectrum, a deep CO ice band is seen toward L1489 IRS,
which we conclude to originate in the disk. The high signal-to-noise
and spectral resolution allowed us to separate it from circumstellar
CO absorption lines, and to do a detailed comparison with laboratory
mixtures. It is, for the first time, found that besides the well known
long wavelength wing of polar H$_2$O:CO ices and the central narrow
absorption peak at 2140 \waven, an additional, separate component is
present on the blue side of the ice band. Both the central peak and
the blue shifted component are due to CO in apolar ices, one due to
pure CO and one due to a mixture of CO$_2$ and perhaps O$_2$ and
N$_2$.  The relatively large depth of the long wavelength, polar wing
relative to the apolar components may suggest that the ices in the
disk of L1489 IRS are thermally processed with respect to the
quiescent Taurus Molecular Cloud material, which is supported by the
low depletion factor ($\sim$7\%) in this line of sight.  This is
likely due to the fact that we see the disk under an angle, and our
pencil absorption beam traces the upper disk layers.

The present work shows that high spectral resolution 4.7 \mum\
observations are a great tool to better understand protostellar disks,
which define the initial conditions of planet and comet formation both
in the solid and gas phase.  This is an exploratory study and needs to
be followed up by observing a larger sample of protostellar disks at
4.7 \mum\ to investigate the influence of parameters such as
foreground contribution, disk inclination, age, and binarity on gas
and ice band profiles.

\acknowledgments

The UCLA NIRSPEC instrument team is acknowledged for constructing an
excellent instrument. We thank M. Brown (Caltech, USA) for providing a
procedure to rectify 2D spectra in IDL, T. Teixeira (University of \AA
rhus, Denmark) for generously providing her compilation of laboratory
CO spectra to us, and the anonymous referee for careful reading of the
manuscript which helped to clarify the description of several aspects
of the analysis.  The research of ACAB at the Caltech Submillimeter
Observatory is funded by the NSF through contract AST-9980846.  The
research of MRH is supported by the Miller Institute for Basic
Research in Science.  NASA support to GAB is gratefully acknowledged.
The authors wish to extend special thanks to those of Hawaiian
ancestry on whose sacred mountain we are privileged to be guests.
Without their generous hospitality, none of the observations presented
herein would have been possible.

\end{document}